\documentclass[onecollarge,fleqn,runningheads]{svjour2}
\smartqed  
\usepackage{graphicx}
\usepackage{mathptmx}      
\usepackage{amssymb}
\usepackage{amsmath}
\usepackage{defn}

\journalname{Theoretical and Computational Fluid Dynamics}
\begin{document}

\title{Secondary Instabilities in Incompressible Axisymmetric
Boundary Layers: Effect of Transverse Curvature 
}

\titlerunning{Effect of transverse curvature on secondary instabilities}        

\author{N. Vinod         \and
        Rama Govindarajan 
}

\authorrunning{Vinod \& Govindarajan} 

\institute{N. Vinod \at
            Engineering Mechanics Unit \\
             Jawaharlal Nehru Centre for Advanced Scientific Research\\
             Bangalore INDIA\\
              \email{nvinod@jncasr.ac.in} \\          
             \emph{Present address:} Dept. of Mechanical Engineering, University of California Santa Barbara  
           \and
           Rama Govindarajan \at
           Engineering Mechanics Unit \\
             Jawaharlal Nehru Centre for Advanced Scientific Research\\
             Bangalore INDIA\\
              \email{rama@jncasr.ac.in} \\          
}

\maketitle

\begin{abstract}
The incompressible boundary layer in the axial flow past
a cylinder has been shown Tutty et. al.(\cite{tutty}) to be stabler than a two-dimensional
boundary layer, with the helical mode being the least stable.
In this paper the secondary instability of this flow is studied.
The laminar flow is shown here to be always stable at high transverse
curvatures to secondary disturbances, which, together with a similar
observation for the linear modes implies that the flow past a thin cylinder
is likely to remain laminar. The azimuthal wavenumber
of the pair of least stable secondary modes ($m_+$ and $m_-$) are related to
that of the primary ($n$) by $m_+=2n$ and $m_-=-n$.
The base flow is shown to be inviscidly stable at any curvature.

\end{abstract}

\section{Introduction}

At low to moderate freestream disturbance levels, the first step in 
the process of transition to turbulence in a \bl is that at some
streamwise location, the laminar flow becomes unstable to linear 
disturbances. While this instability and the events that follow
have been investigated in great detail for two-dimensional boundary layers 
during the past century, much less work has been done on its
axisymmetric counterpart, the incompressible \bl in the flow past a cylinder,
notable exceptions being the early and approximate linear stability analysis 
of Rao\cite{gnv} and the more recent and accurate one of Tutty et. al.\cite{tutty}.
In Rao's work, the equations were not solved 
directly and the stability estimates had severe limitations. 
Tutty et. al.\cite{tutty} showed that 
non-axisymmetric modes are less stable than axisymmetric ones.
The critical Reynolds number was found to be $1060$ for
the $n=1$ mode and $12439$ for $n=0$. The instability 
is thus of a different character from that in two-dimensional
boundary layers, since Squire's (1933) theorem, stating that the
first instabilities are two-dimensional, is not applicable in this case.
The expected next stage of the process of transition
to turbulence, namely the secondary modes of instability, have not been
studied before, to our knowledge, although the
turbulent flow over a long thin cylinder has been studied by
Tutty \cite{tutty08}, who computed the meanflow quantities to within experimental 
accuracy.
Practical applications, on the other hand, are numerous. For example,
the axial extent of turbulent flow would determine the 
signature that submarines leave behind themselves, apart from the drag. 
The transition to turbulence over the bodies of large fish
too would be partially controlled by transverse curvature. 

The secondary instability of incompressible laminar flow past
an axisymmetric body is thus the focus of this paper. We present results
only for the flow past a cylinder, but the equations derived here
and the solution method can be used for arbitrary axisymmetric bodies.
We show that the overall effect of transverse curvature on
incompressible boundary layers is to stabilise  
secondary disturbances. Remarkably no instability is found at any
Reynolds number at higher curvatures, i.e., when the boundary layer
thickness becomes comparable to the body radius. This implies that the
boundary layer past a thin cylinder would tend to remain laminar, or to
relaminarise downstream even were it to go turbulent.

We note in contrast that longitudinal curvature, and the resulting 
G\"{o}rtler (1940) vortices on concave walls, have been well studied 
(see e.g. Benmalek and Saric \cite{saric94}) and so have instabilities in
three-dimensional boundary layers due to streamline curvature
\cite{itoh96}.

\section{Mean flow}
\label{chp_mean}

The unperturbed laminar flow is obtained by solving the incompressible 
steady boundary layer equation for the axial component of velocity:
\begin{figure}
\begin{center}
\includegraphics[scale=0.5]{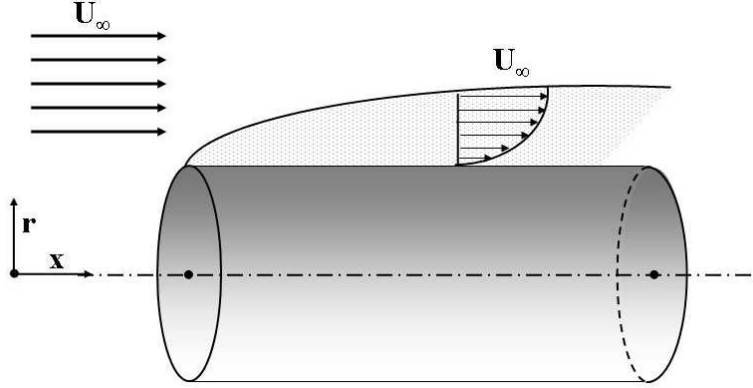}
\vskip5mm   
\caption{ Schematic diagram showing the coordinate system.}
\label{fig:cyl}
\end{center}
\vskip5mm   
\end{figure}
\be
U\popa U x + V \popa U r = \frac{1}{R} \Big(\popasq U r + {1 \over r}\popa U r 
\Big), 
\label{eq:mean}
\ee
together with the continuity equation
\be
\popa U x + \popa V r + \frac V r=0,
\label{eqn:cont}
\ee
and the boundary conditions
\be
U(0,r)=1, \qquad 
U(x,1)=0, \qquad V(x,1)= 0 \qquad {\rm and} \qquad U(x,\infty)=1.
\label{eqn:bc}
\ee
Here the 
$x$ coordinate is along the surface of the body and $r$ is normal to the 
body surface and measured from the body axis. The respective 
velocity components in these co-ordinates are $U$ and $V$.
The length  and velocity scales which have been used for 
non-dimensionalisation
are the body radius, $r_0$, and the freestream velocity, $U_{\infty}$, 
respectively, so the \re is
\be
R\equiv\frac{U_{\infty} r_0}{\nu}.
\ee
The solution is obtained by a 3-level implicit finite difference scheme 
on a uniform grid. At the leading edge, two levels of initial data are provided,
and downstream marching is commenced. The discretised equation is second order 
accurate in $\Delta x$ and $\Delta r$, and is unconditionally stable in the von 
Neumann sense. A fairly fine grid in the $r$ direction is necessary to capture 
the velocity and its derivatives accurately. With a grid size of $10^{-3}$ in 
the $x$ and $r$ directions the results are accurate up to 7 decimal places.

\begin{figure}
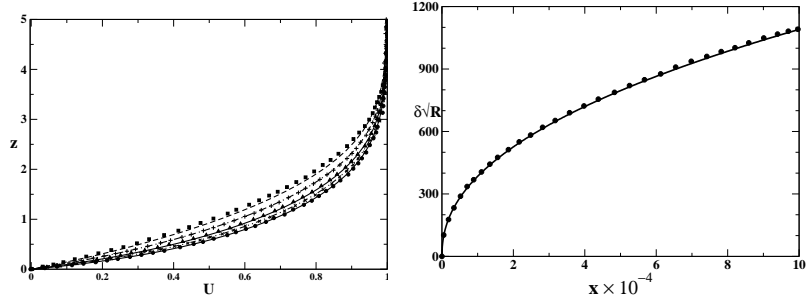

\begin{center}
\includegraphics[scale=0.22]{comp_ut}
\hskip2mm
\includegraphics[scale=0.22]{delta}
\vskip5mm
\caption{(a) Velocity profiles compared to those of Tutty et. al.\cite{tutty} at 
$R=10^4$. The ordinate gives $z=\sqrt{R/x^*}(r-1)$. The lowest curve is 
at $x=10^5$ and the topmost curve is at $x=398$. The intermediate curves 
are spaced at intervals of $x^{1/2}$. (b) Dimensionless 
boundary layer thickness $\delta\sqrt{R}$ at $R=10^4$. In both figures,
the symbols are from \cite{tutty}, while the lines are present results. }
\label{tutty_u}
\end{center}
\end{figure}

\begin{figure}
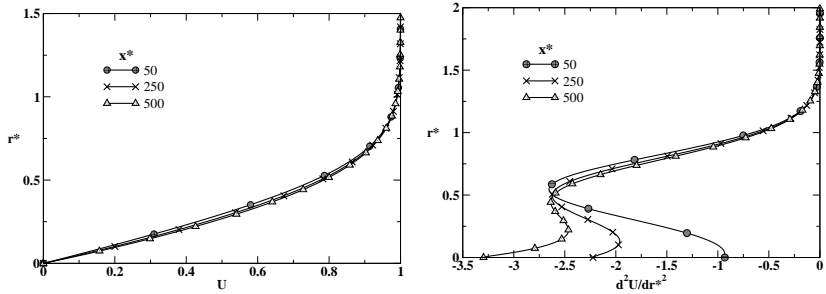

\begin{center}
\vskip0.1in \includegraphics[scale=0.22]{u4k.eps}
\hskip2mm   
\includegraphics[scale=0.22]{udd4k}
\vskip0.1in 
\caption{(a) Velocity profile at different streamwise locations, at a 
Reynolds number of $4000$. The coordinate $r^* \equiv (r-1)/\delta$. (b) 
Second derivative of streamwise velocity, $d^2 U/d r^2$. }
\label{u_ns_R4}
\end{center}
\end{figure}

Velocity profiles at a Reynolds number of $10000$ are seen in figure 
\ref{tutty_u}(a) to be in good agreement with the results of Tutty et. al.\cite{tutty}. The 
dimensionless boundary layer 
thickness $\delta$ ($\equiv r_{0.99}-1$, where $U_{r_{0.99}}=0.99$) is 
plotted in figure \ref{tutty_u}(b).
When scaled by the local boundary layer thickness, there is not much of a 
difference visible in the profiles, as seen in figure \ref{u_ns_R4}(a) where
the \re is $4000$.
Here the coordinate $r^*$ is measured from the body surface. A marked 
difference near the wall is however evident in the second derivative of 
the velocity (figure \ref{u_ns_R4}(b)). This difference is seen below to 
significantly affect stability behaviour. Specifically, an increasingly 
negative second derivative is indicative of a fuller, and therefore 
more stable, profile downstream.

The boundary layer does not obey similarity, since there are two parameters 
in the problem, $R$ and the surface curvature $S_0$ defined below. 
Defining
\be
S_x\equiv (4\nu x_d/U_\infty r_0^2)^{1/2},
\ee
where the subscript $d$ denotes a dimensional quantity, 
we may convert the partial differential equation (\ref{eq:mean}) to an 
ordinary differential equation in the variable $\chi=r^2$ ($r$ is 
 non-dimensionalised with body radius $r_0$.)
\be
\chi g'''+(1+  {1 \over 2S_x^2}g)g''=0.
\label{eq:sim}
\ee
Here $g'=2U$, and the explicit dependence on $x$ is contained in 
$S_x$. It is evident that the velocity profiles would be self-similar were the
quantity $x/R$ to be held constant. The momentum thickness in an axisymmetric 
boundary layer is of the form
\be
\theta=-r_0+\sqrt{r_0^2+2I}, \qquad
{\rm where}
\qquad
I\equiv\int_{r_0}^{\infty}U (1- U) r_d dr_d.
\ee
The displacement 
thickness may be similarly defined. The surface curvature, i.e., the ratio of a 
typical boundary layer thickness to the body radius is defined here as
\be
S_0 \equiv \frac{\theta}{r_0}.
\ee

\section{Linear stability analysis}
\label{chp_stab}

Based on present wisdom, and our own experience in boundary layer flows,
we make the assumption that non-parallel effects are small. The equations in
this section are the same as those of Tutty et. al. \cite{tutty}, expressed in terms of the
variables introduced by Rao (1967). Flow 
quantities are decomposed into their mean and a fluctuating part, e.g.
\beq
\vec{v_{tot}}&=&U(r)\vec{x}+\vec{v}(x,r,\gamma,t)
\eeq
where $\vec{v} = u\vec x + v \vec r + w \vec \gamma$, $\gamma$ being the 
azimuthal coordinate. Disturbance velocities are expressed in terms of 
generalized stream-functions $\psi$ and $\phi$ as 
\be
u={1 \over r} {\popa \psi r},  \;
v=-{1 \over r}\left(\popa \psi x + \popa \phi \gamma \right) \; {\rm and } \;
w={\popa \phi r} .
\label{eq:phi}
\ee
In normal mode form
\be
(\psi,\phi)(x,r,\gamma)=\frac{1}{2}\Big((\Psi,\Phi)(r) \exp[\i(\alpha 
x+n\gamma -\omega t)] + {\rm c.c}\Big).
\ee
\label{eqn:psi}
Here $\Phi (r)$ and $\Psi (r)$ are the amplitudes of the disturbance 
stream-functions, $\alpha$ is the wave number in the streamwise direction 
and $n$ is the number of waves encircling the cylinder. The value of $n$ 
is positive or negative for anti-clockwise or clockwise wave propagation
respectively. In the temporal stability analysis carried out here, the 
imaginary part of the frequency $\omega$ gives the growth rate of the 
disturbance.

Linearising the Navier-Stokes for small disturbances and eliminating 
the disturbance pressure results in two fourth-order ordinary differential 
equations in $\Psi$ and $\Phi$, given by

\begin{equation}
\centering
\oma \left(\Psi'' - S\Psi'  - \alpha^2\Psi\right) - \Psi
(U'' - SU' ) - \oma\alpha n \P-{n \over \alpha}\left(U''\P+U'\P'-
SU'\P\right)) \nonumber
\end{equation}
\begin{equation}
\centering
={1 \over i\alpha R} \bigg[\Psi^{iv} -2 S \Psi''' + 3 S^2 \Psi''- 3 S^3 \Psi'
- 2\alpha^2 (\Psi''- S \Psi') + \alpha^4\Psi \nonumber
\end{equation}
\begin{equation}
\centering
-n^2S^2\left(\Psi'' - 3S\Psi'  - \alpha^2\Psi\right)
-n\alpha[\P''+S\P'-(\alpha^2+n^2S^2)\P]\bigg]
\label{dist1}
\end{equation}
and  
\begin{equation}
\oma(\P''+S\P'-n^2S^2\P)+U'\P'-\oma n\alpha S^2\Psi
={1 \over i\alpha R}\bigg[\P^{iv}+2S\P''' \nonumber
\end{equation}
\begin{equation}
-S^2\P''(1+2n^2) -\alpha^2\P''+ S^3(1+2n^2)\P'-\alpha^2S\P'
-S^4(4n^2-n^4)\P+\alpha^2 n^2S^2\P \nonumber
\end{equation}
\begin{equation}
-n\alpha S^2\Psi''+
3\alpha nS^3 \Psi'-(4\alpha n S^4-\alpha^3nS^2-\alpha n^3S^4)\Psi\bigg].
\label{dist2}
\end{equation}
Here $S=\theta/r$, and the boundary conditions are
\begin{equation}
\Psi = \Psi' = 0, \quad \quad \Phi = \Phi' = 0 \quad {\rm at} \quad r=1, 
\label{wallbc}
\end{equation}
and
\begin{equation}
 \Psi = \Psi' \to 0, \quad \quad \Phi = \Phi' \to 0 
\quad {\rm as} \quad r \to \infty.
\label{outbc}
\end{equation}
Upon putting $S=0$ and letting $n\to\infty$ such that $nS$ tends to a 
finite quantity corresponding to the spanwise wavenumber, $\beta$, 
equations \ref{dist1} and \ref{dist2} reduce with some algebra to the 
three-dimensional Orr-Sommerfeld and Squire's equations for \bls 
on \2d surfaces (see e.g. \cite{hennibook}). 

The rates of production $W_+(r)$ and dissipation $W_-(r)$ of disturbance 
kinetic energy are given by
\be
W_+(r)=-{1 \over 2}\Big(vu^* + v^*u \Big){dU \over dr},
\label{eqn:prod}
\ee
and,
\begin{equation}
W_-(r)={1 \over Re}\Big(\alpha^2 (u u^*+v v^* +w w^*)
+u' u'^*+v' v'^*+w' w'^* \nonumber  
\ee
\be
+{1 \over r^2}\big[n^2u u^* + (1+n^2)(v v^*+w w^*)
+2in(v^*w-vw^*)\big] \Big)
\label{eq:dsp}
\end{equation}
where the superscript * denotes the complex conjugate.
Note that the last term in \ref{eq:dsp} is derived from squares of
magnitudes, and is thus real and positive. 

\subsection{Inviscid stability characteristics}
\label{sec:invisc}
It is instructive to first study what happens under inviscid conditions.
For two-dimensional flow, the existence of a point of inflexion in the 
velocity profile is a necessary condition \cite{rayl} for inviscid 
instability. The axisymmetric analog of this criterion has been derived
for various situations e.g., Duck \cite{duck90} obtained the 
generalised criterion for axisymmetric disturbances on axisymmetric 
compressible boundary layers.

In brief, in the inviscid limit we may eliminate all variables except 
$v$ in the momentum and continuity equations for the linear perturbations, to get

\begin{equation}
(U-c) \left [ v^{\prime \prime} + {{\left ( 3 n^2 + \alpha^2 r^2 \right ) }
\over {\left ( \alpha^2 r^2 + n^2 \right ) }}{v^\prime \over r} - {{\left (
\alpha^2 r^2 + n^2 + 2 \right ) } \over {\left ( \alpha^2 r^2 + n^2 \right)}}
\alpha^2 v + (1-n^2) {v \over r^2} \right ]
 - \left [ U^{\prime \prime} -
{{\left ( \alpha^2 r^2 -n^2  \right ) } \over r {\left ( \alpha^2 r^2 +
n^2 \right ) }} U^\prime \right] v = 0.
\label{inviscid0}
\end{equation}
From a procedure similar to that for two-dimensional flows, a necessary 
condition for instability, that the quantity $I \equiv   
U ^{\prime \prime} - [{{(\alpha^2r^2 - n^2)} / r / {(\alpha^2r^2 + n^2)}}] 
U ^\prime $, has to change sign somewhere in the domain, is obtained.
Letting $r \to \infty$, we recover the two-dimensional Rayleigh criterion.

Unlike in two-dimensional flows, the quantity $I$ 
depends on the streamwise and azimuthal wavenumbers, but in order to check
for instability it is sufficient to evaluate the limiting cases
$I_1$ and $I_2$ respectively for $\alpha/n\to 0$ and $n/\alpha\to 0$. 
Using equations \ref{eq:mean} and \ref{eq:sim}, $I_1$ and  $I_2$ can be 
written as 
\beq
I_1&=&U''-\frac{U'} r =  \frac{r^2}{8x_d^2}g''',\\
 {\rm and } \quad
I_2&=&U'' + \frac{U'} r =  - \frac{1}{4R x_d/r_0} g g''.
\eeq

\begin{figure}
\begin{center}
\vskip0.2in\includegraphics[scale=0.35]{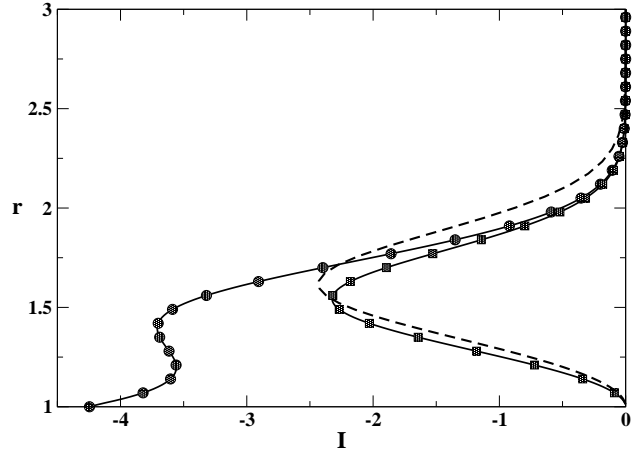}
\caption{Sample plot of $I_1$ (circles) and $I_2$ (squares) at a \re of 
$5000$ and a curvature of $S_0=0.8$.
The $U''$ of the Blasius profile is shown by the dotted line.
}
\label{fig:sup}
\end{center}
\end{figure}

At the wall and at the freestream, $g$ and $g''$ are equal to 
zero, so $I_2$ is zero too. In between, $I_2$ is always negative since both 
$g$ and $g''$ are positive. $I_1$ is negative everywhere as well,
i.e. $I$ never changes sign. 
In figure \ref{fig:sup} these quantities are plotted for a sample case
($S_0=0.8$).
We conclude that the incompressible \ax laminar \bl on a circular 
cylinder is inviscidly stable to axisymmetric and non-axisymmetric
disturbances at any curvature.

In two-dimensional boundary-layers, the inflexion point criterion has
provided a general guideline for viscous flows as well, since a flow with 
a fuller velocity profile typically remains stable up to a much higher 
Reynolds number. We may therefore expect from figure \ref{fig:sup} that an 
axisymmetric 
boundary layer will be more stable than a two-dimensional one. Also as the 
curvature increases (not shown) the tendency to stabilise will be higher.
Note that a change of sign in $I$ may occur on converging bodies. We do 
not consider 
that case here, but mention that the axisymmetric analog of Fjortoft's 
theorem,
\begin{equation}
(U-U_s) \left [ U ^{\prime \prime} - {{(\alpha^2r^2 - n^2)}
\over r {(\alpha^2r^2 + n^2)}} U ^\prime \right ] \le 0,
\label{fjqrtoft}
\end{equation}
where $U_s$ is the velocity at the inflection point, being a stricter 
criterion than the Rayleigh could then be used. The above may easily be
obtained again by a procedure similar to that in two dimensions.

\subsection{Numerical method and validation}
\label{linstab}

Equations \ref{dist1} to \ref{outbc} form an eigenvalue problem, which is 
solved by a Chebyshev spectral collocation method. The transformation
\be
y(i)=1+{\Big(1+y_C(i)\Big)\kappa \over 1 + {2\kappa \over L} - y_C(i)},
\label{grid}
\ee
where
\be
y_C(i)=\cos({\pi i \over N}), \; \; \;  \; \; \; i=0,1...,N.
\ee 
are the collocation points, is used to obtain a computational domain extending 
from $r=1$ to $r=L+1$ and to cluster a larger number of grid points close to 
the wall, by a suitable choice of $\kappa$. We ensure that $L$ is at least $5$ 
times the \bl thickness, so that the far-field boundary conditions are 
applicable. Eigenvalues obtained using $81$ and $161$ grid points are identical 
up to the sixth decimal place.

\begin{table*}
\begin{center}
\begin{tabular}{|c|rrcc|rrcc|}
\hline
&\multicolumn{4}{c|}{Tutty et. al. \cite{tutty}}& \multicolumn{4}{c|}{Present} \\
$n$ & $x_c$ & $R_c$ & $\alpha_c$ & $c_r$ & $x_c$ & $R_c$ & $\alpha_c$ & $c_r$\\ 
\hline
0 & 47.0 & 12439 & 2.730 & 0.317 & 47.0& 12463& 2.730&  0.318 \\
1 & 543.0 & 1060 &0.125 & 0.552 & 581.0 & 1013 & 0.115 &0.552 \\
2 &91.1 & 6070 &  0.775 & 0.442 & 91.0& 6093& 0.775&0.421 \\
3 & 43.4 & 10102 & 1.600 & 0.403 & 43.0& 10110& 1.580& 0.410 \\
4 & 26.8 & 13735 & 2.540 & 0.398 & 27.0& 13742& 2.520& 0.401\\
\hline
\end{tabular}
\caption{Critical Reynolds number and other parameters for different modes,
in comparison with \cite{tutty}. The streamwise location where instability 
first occurs is denoted as $x_c$. $\alpha _c$ and $c_r$ are the 
streamwise wavenumber and phase speed corresponding to the critical 
Reynolds number $R_c$.}
\label{tab:comp}
\end{center}
\end{table*}

We compare our critical values with those of Tutty et. al. \cite{tutty} in table 
\ref{tab:comp}, and find them to be in reasonable agreement. The helical 
mode ($n=1$) is destabilised first at a \re of $1013$, and $x=581$. 
The axisymmetric ($n=0$) mode is unstable only above a \re of $12463$. 

\begin{figure}
\begin{center}
\includegraphics[scale=0.40]{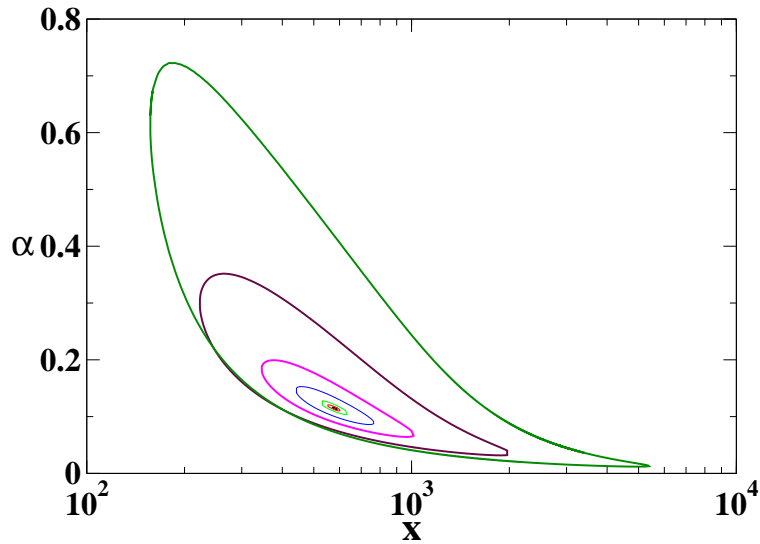}
\vskip0.1in \caption{Neutral stability loops for the non-axisymmetric mode
$n=1$ along the axial coordinate. The Reynolds number for the innermost curve is 1014 and that of
the outermost curve is 5000. The Reynolds numbers for the other curves from
inside to outside are 1015,1020,1060,1200 and 2000 respectively.}
\label{neu1}
\end{center}
\end{figure}

\begin{figure}
\begin{center}
\vskip0.2in \includegraphics[scale=0.35]{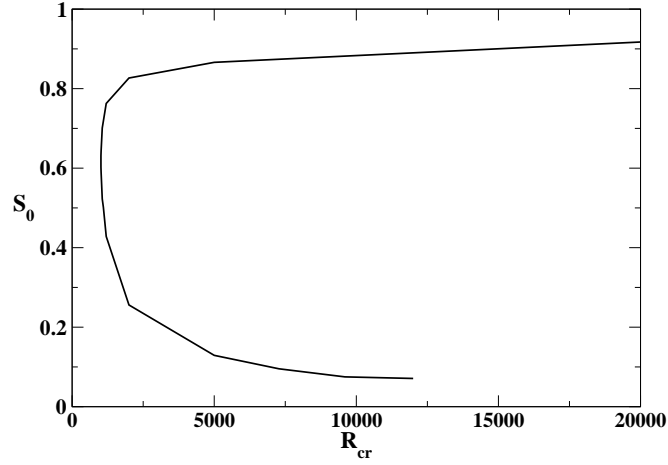}
\vskip0.1in \caption{Critical \re as a function of curvature for the mode 
$n=1$.}
\label{rcr}
\end{center}
\end{figure}

It is only the helical ($n=1$) mode which is unstable over a significant axial 
extent of the cylinder. Even this mode is never unstable for curvatures above 
$S_0=1$, as may be seen in figure \ref{rcr}. At curvature levels below this, 
as well as at low Reynolds numbers, the helical mode is expected to decide 
dynamics, since the ranges of instability of other modes are subsets of the range of 
the $n=1$ mode.

\begin{figure}
\begin{center}
\includegraphics[scale=0.35]{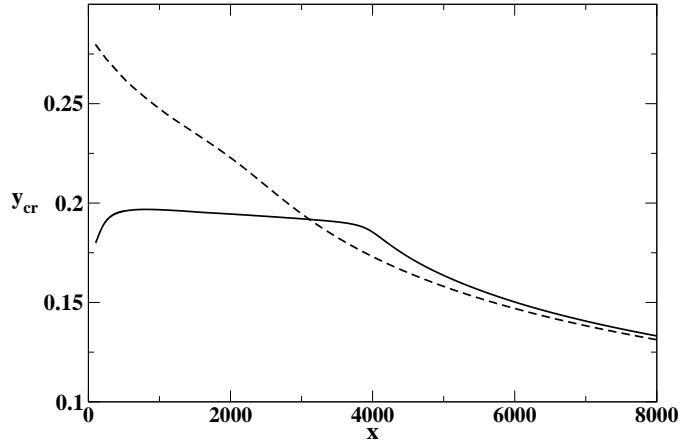}
\vskip0.1in \caption{Height of the critical layer as a function
of the streamwise distance, $n=1$, $\alpha=0.125$. Dashed line: 
$Re=2000$; solid line $Re=10,000$.}
\label{phase}
\end{center}
\end{figure}

Figure \ref{phase}(a) shows the downstream variation of the critical 
layer location $y_{cr}$, where $U(y_{cr})=c_r$. It is seen that as one 
moves downstream, i.e., as the curvature increases for a given Reynolds 
number, the critical layer moves closer to the wall. Since the production 
layer scales as $2\pi (y_{cr}+1)r_0$, as the cylinder becomes thinner and thinner,
the cross-sectional area over which production is possible is much smaller, 
explaining the stabilization at large curvature. The energy budget shows that the
production layers of several unstable modes overlap, and this could give rise to 
interactions between them, so the nonlinear stability could be very different from
that in a planar boundary layer.

\section{Secondary instabilities}
\label{chp_sec}

A laminar flow containing linear disturbances of a significant 
amplitude is unstable to secondary modes.
The $\Lambda$-structures seen in  \cite{kleb}
and Kachanov \cite{kach94}, considered to be the precursors of turbulent spots, are 
a signature of these modes. As a rule of thumb, nonlinearity in boundary 
layers becomes 
detectable when the amplitude of the linear (primary) disturbance is $1\%$ 
of the mean flow. 

The approach we follow is standard, as in \cite{88herb}. The periodic basic 
flow is expressed as,
\be
\vec{v_{basic}}=\vec {U(r)}+ A_p \vec{v_p}
\label{eq:form}
\ee
where we have introduced a subscript $p$ for the primary (linear) disturbance.
$A_p$ is the ratio of the amplitude of disturbance velocity to the freestream 
velocity. $v_p$ is the disturbance velocity of primary modes, obtained from
linear stability analysis. The secondary disturbance, in normal mode form, is 
\be
{\vec{{v_s}}}=\frac{1}{2}\Big(\vec{v}_{+}(r)
\exp[i(k_+ x+ m_+\theta -\omega_+ t)] + 
\vec{v}_{-}(r) \exp[i(k_- x+ m_-\theta -\omega_- t)] + {\rm c.c}
\Big),
\label{eq:secv}
\ee
where
$k_+$ and $ +k_-$ are the streamwise wavenumbers of the secondary waves. 
The azimuthal wavenumbers of secondary waves are $m_+$ and $m_-$. 

Equations for the secondary instability are obtained by substituting
\ref{eq:form} and \ref{eq:secv} into the Navier-Stokes equations,
retaining  linear terms in the secondary disturbances, and deducting the
primary stability equations.
The streamwise component of the velocity $u_+$ and $u_-$ are eliminated
using the continuity equation. The final equations contain
$v$, $w$  and $p$.
On averaging over
$x$, $\theta$ and $t$, only the resonant modes survive, which are related as
follows:
\be
k_+ + k_- = \alpha, \qquad m_++m_-=n \qquad {\rm and} \qquad  (\omega_{+} + \omega_-)_r =
\omega.
\label{eq:azim}
\ee

The final secondary instability equations are

$$
\Big\{(U-c)(S +D)-U'-\frac{i}{k_+R}\Big[S\Big(k_+^2+(m_+^2-1)S^2\Big)
+ \Big(k_+^2+({m_+^2+1})S^2\Big)D- 2S D^2 -D^3\Big]\Big\}v_{+}
$$
$$
+\Big\{(U-c)im_+S+\frac{m_+S}{k_+R}\Big(k_+^2+(m_+^2-1)S^2 +S D 
-D^2\Big)\Big\}w_{+} -ik_+p_{+}
$$
$$
+\frac{1}{2k_-}\Big[k_+Su_p+iS^2v_p-m_-S^2w_p+k_-u_p'+(k_+u_p-iSv_p-m_-Sw_p)D-iv_pD^2\Big]v_{-}^*
$$
\be
+\Big[\frac{iS}{2}\Big(n-\frac{k_+}{k_-}m_-\Big)u_p+\frac{m_-S^2}{2k_-}v_p
+\frac{im_-^2S^2}{2k_-}w_p-\frac{m_-S}{2k_-}v_pD\Big]w_{-}^*=0
\label{eq:sec1}
\ee

$$
-\frac{2im_+S^2}{R}v_{+}+\Big[ik_+(U-c) + \frac 1 R\Big(k_+^2+(m_+^2+1)S^2
-S D- D^2\Big)\Big]w_{+} + im_+Sp_{+}
$$
\be
+\Big[\frac{S}{2}\Big(\frac{\alpha}{k_-}-1\Big)w_p+\frac 1 2 w_p' +\frac{\alpha}{2k_-}w_pD\Big]v_{-}^*
+\Big[-\frac{ik_-}{2}u_p+\frac{S}{2}v_p+\frac{iS}{2}\Big(m_+-\frac{m_-\alpha}{k_-}\Big)w_p\Big]w_{-}^*=0
\label{eq:sec2}
\ee
$$
\Big[ik_+(U-c) + \frac 1 R\Big(k_+^2+(m_+^2+1){S^2}
-SD- D^2\Big)\Big]v_{+}+\frac{2im_+S^2}{R}w_{+} + p_{+}'
+\Big[-\frac{ik_-}{2}u_p
$$
\be
-\frac{im_-S}{2}w_p+\frac{\alpha S}{2k_-}v_p+\frac{v_p'}{2}
+\frac 1 2 \Big(1+\frac{\alpha}{k_-}\Big)v_pD\Big]v_{-}^*
+\Big[\frac{iS}{2}\Big(n-\frac{\alpha m_-}{k_-}\Big)v_p-\frac{Sw_p}{2}\Big]w_{-}^*=0
\label{eq:sec3}
\ee
with three corresponding equations in $v_{-}^*$, $w_{-}^*$  and $p_{-}^*$.
The operator $D$ stands for differentiation with respect to the radial coordinate.

The boundary conditions are
\be
 \vec{v_s} =0 \quad {\rm at} \quad r = 1, \qquad
\vec{v_s} \rightarrow 0  \quad {\rm as} \quad r \rightarrow \infty,
\qquad {\rm and} \quad p \rightarrow 0 \quad  {\rm as} \quad r 
\rightarrow \infty. 
\ee
\label{bcs}

For the flow under consideration, the growth/decay rates of primary modes are
small, hence $d (A_p)/dt$ can be neglected during one period of time.
As mentioned above, Squires theorem does not apply
and therefore the primary modes must be taken to be three-dimensional.
Equations  \ref{eq:sec1} to  \ref{eq:sec3} reduce to
the secondary instability equations of a flat plate boundary layer
by letting $S=0$, $m_+S \to \beta$, $m_-S \to -\beta$ and $n_S = 0$. 
The system is solved as before. Disturbance growth rates for 
a zero pressure gradient boundary layer agree well
with those of \cite{88herb}.

\subsection{Results}

\begin{figure}
\begin{center}
\includegraphics[scale=0.27]{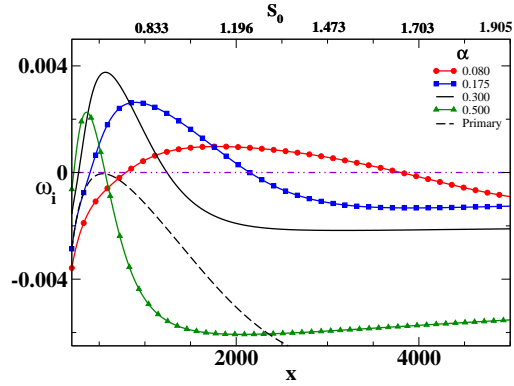}
\vskip0.1in \caption{Growth rate of secondary disturbance waves along the 
cylinder axis for the non-axisymmetric mode $n=1$,  $m_-=-1$ with $A_p=0.02$ at 
$R=1000$. The most unstable primary mode is shown by the dashed line.  } 
\label{n1_1K}
\end{center}
\end{figure}

The main finding is that for high levels of curvature
the flow is stable to secondary modes (as well as the linear modes), but 
secondary modes 
can extend the curvature range over which disturbance growth is possible.
As in the case of a \2d \bl
subharmonic modes are dominant here too. 
The axial wavenumber of the least stable secondary mode is $k_+= k_- =\alpha/2$.
At other values $k_+$ or $k_-$, the growth is smaller. This is similar to the
behaviour in flat plate  boundary layers.
The most unstable 
secondary modes are of opposite obliqueness, with
azimuthal wavenumber $m_+=2n$ and $m_-=-n$.

is presented in figure \ref{n1_1K}a. The amplitude $A_p$ of the primary 
wave is taken to be $2\%$ of $U_\infty$, but the answers do not depend
qualitatively on this choice. The flow is seen to be unstable to 
secondary modes under conditions where all primary disturbances decay.
For comparison the growth rate of the least stable primary disturbance
($\alpha=0.125$ and $n=1$) is shown as a dotted line.
At small primary wave numbers the growth rate is found to be small
but for a wider band of streamwise locations. As the wavenumber increases the growth rate
also increases and instability is restricted to a narrow streamwise region.
The maximum growth occurs when $\alpha=0.30$, and $k_+=k_-=0.15$. As discussed 
earlier $m_+=2$ and $m_-=-1$ in this case.
It is found that the secondary modes also decay at higher curvature,
but the extent of curvatures for which the instability is sustained is much larger.

\begin{figure}
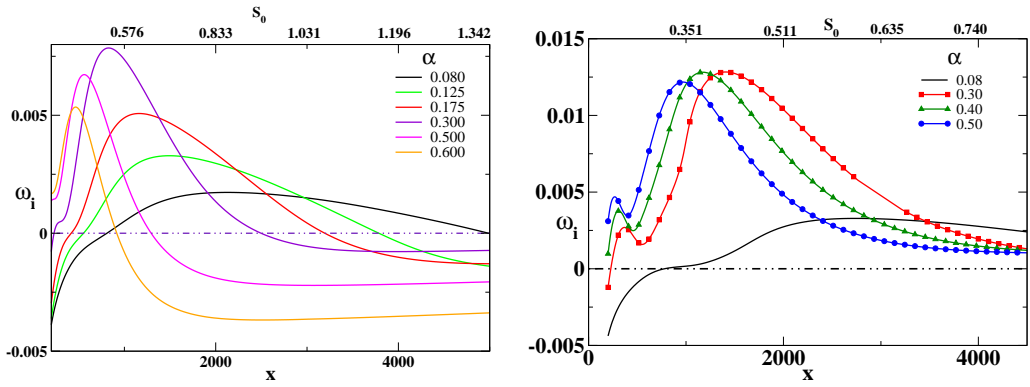

\begin{center}
\includegraphics[scale=0.27]{n1R2K.eps}
\hskip4mm
\includegraphics[scale=0.27]{n1R5K.eps}
\vskip0.1in \caption{Growth rate of secondary disturbance along the cylinder axis
for (a) $R=2000$ and (b) $R=5000$. The other parameters are the same as in figure \ref{n1_1K}. }
\label{n1_2K}
\end{center}
\end{figure}

In figure \ref{n1_2K} the growth of secondary waves at higher Reynolds numbers
are plotted. The flow conditions are same as in figure \ref{n1_1K} except the Reynolds numbers.
In \ref{n1_2K}a  \re is 2000 while   \ref{n1_2K}b is for \re 5000.
The primary modes amplify substantially in these cases.
The behaviour at a higher Reynolds number, as seen in 
figure \ref{n1_2K}, is as expected. 

Incidentally a small growth is found at small curvatures (figure \ref{n1_2K}b) at this \re.
However the this growth is very smaller compared to that at higher locations.

\begin{figure}
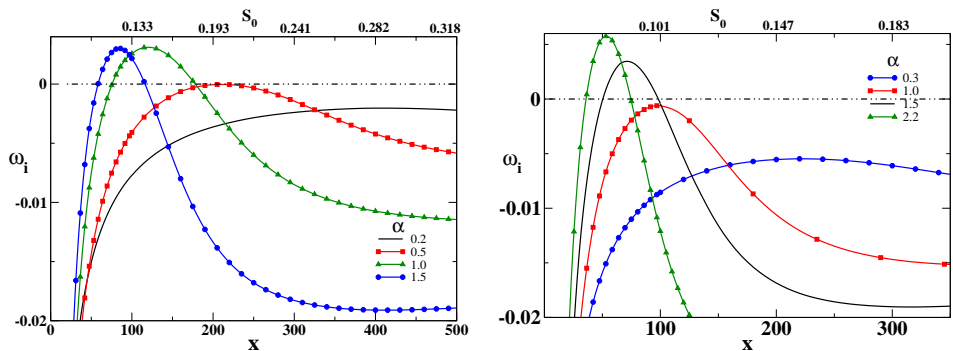

\begin{center}
\includegraphics[scale=0.25]{n2s3K.eps}
\hskip3mm
\includegraphics[scale=0.25]{N3_5Kn.eps}
\vskip0.1in \caption{Growth rate of secondary disturbance modes with
$A_p=0.02$. (a) $n=2$, $m_+=4$ and $m_-=-2$ at $R=3000$.
(b) $n=3$, $m_+=6$ and $m_-=-3$ at $R=5000$.}
\label{n2_3K}
\end{center}
\end{figure}

The least stable secondary modes for other values of the
azimuthal wavenumber $n$ are shown in figure \ref{n2_3K}. It is clear that
in the range of Reynolds numbers of interest, these modes are not expected to 
dominate.
The axisymmetric mode is not shown, but does not afford any surprise either.
We have computed growth rates for different modes ranging $n=1$ to $n=4$, but have shown only 
the results for $n=1$ and $n=2$. There is no qualitative difference in  other two modes,
where the relationship of $m_+$ to $n$ holds good.

\section{Conclusions}

The boundary layer in the flow past a cylinder is stable 
to (linear and) secondary disturbances at curvatures higher than $S_0 \sim
O(1)$, i.e. when the radius of the body is of the order of or less than
the local boundary layer thickness (shown here in terms of momentum
thickness).
This indicates that a turbulent axisymmetric boundary layer, especially
over a thin body, could have a tendency to relaminarise downstream. 
The flow is inviscidly stable at any curvature.
Transverse curvature thus has an overall 
stabilising effect, acting via the mean flow and directly through 
the stability equations. 

The production layers of the disturbance kinetic 
energy of these modes have a significant overlap, which gives rise
to the possibility of earlier development of nonlinearities. 
Thus,
while transverse curvature delays
the first instability, it can contribute once instability 
sets in to a quicker and different route to turbulence. 

Secondary disturbances remain unstable at larger curvatures than linear 
modes.
However there is again a maximum curvature ($S_0 \approx 2$ for $A_p=0.02$)
above which all disturbances decay. The most unstable secondary modes 
are always those whose azimuthal wavenumbers are related to that of the 
primary mode by $m_+=2n$ and $m_-=-n$. For a helical primary mode (which is the
most unstable) this means
that one of the secondary perturbations is helical as well, but of opposite 
sense, while the other has the same sense but two waves straddle the body.
We contrast this to a Blasius boundary layer, where the most unstable secondary 
mode is three-dimensional while the most unstable primary is two-dimensional.
There, the spanwise wavenumber $\beta$ is of the same order as the 
streamwise wavenumber, and two sets of identical looking waves travel in the 
positive and negative spanwise directions. We do not yet have an explanation for
our observation, except to say that in a coordinate moving with the primary
wave, the observed most unstable secondary is the simplest combination
containing one forward propagating (in the azimuthal direction) and one 
backward propagating wave relative to the primary.
As for the axial wavenumber, the subharmonic modes are least stable,
as in two-dimensional boundary layers. 

Our studies indicate that experimental and numerical studies of this
problem could uncover new physics about the transition to turbulence.
We hazard a prediction that this process will be significantly different from
two-dimensional flow.


\begin{thebibliography}{10}
\providecommand{\url}[1]{{#1}}
\providecommand{\urlprefix}{URL }
\expandafter\ifx\csname urlstyle\endcsname\relax
  \providecommand{\doi}[1]{DOI~\discretionary{}{}{}#1}\else
  \providecommand{\doi}{DOI~\discretionary{}{}{}\begingroup
  \urlstyle{rm}\Url}\fi

\bibitem{saric94}
Benmalek, A., Saric, W.S.: Effects of curvature variations on the nonlinear
  evolution of {G}\"{o}rtler vortices.
\newblock Phys Fluids \textbf{6}(10), 3353--3367 (1994)

\bibitem{duck90}
Duck, P.W.: The inviscid axisymmetric stability of the supersonic flow along a
  circular cylinder.
\newblock \jfm \textbf{214}, 611--637 (1990)

\bibitem{88herb}
Herbert, T.: Secondary instability of boundary layers.
\newblock Annu. Rev. Fluid Mech. \textbf{20}, 487--526 (1988)

\bibitem{itoh96}
Itoh, N.: Simple cases of the streamline-curvature instability in
  three-dimensional boundary layers.
\newblock J. Fluid Mech. \textbf{317}, 129--154 (1996)

\bibitem{kach94}
Kachanov, Y.S.: Physical mechanisms of laminar-boundary layer transition.
\newblock Annu. Rev. Fluid Mech. \textbf{26}, 411--482 (1994)

\bibitem{kleb}
Klebanoff, P.S., Tidstorm, K.D., Sargent, L.M.: The three-dimensional nature of
  boundary layer instability.
\newblock J. Fluid. Mech. \textbf{12}, 1--34 (1962)

\bibitem{gnv}
Rao, G.N.V.: Effects of convex transverse surface curvature on transition and
  other properties of the incompressible boundary layer.
\newblock Ph.D. thesis, Dept. of Aerospace Engg., Indian Institute of Science
  (1967)

\bibitem{rayl}
Rayleigh: On the stability of certain fluid motions.
\newblock Proc. Math. Soc. Lond. \textbf{11}, 57--70 (1880)

\bibitem{hennibook}
Schmid, P.J., Henningson, D.S.: Stability and transition in shear flows.
\newblock Springer-Verlag, New York (2001)

\bibitem{tutty08}
Tutty, O.R.: Flow along a long thin cylinder.
\newblock J. Fluid Mech. \textbf{602}, 1--37 (2008)

\bibitem{tutty}
Tutty, O.R., Price, W.G., Parsons, A.T.: Boundary layer flow on a long thin
  cylinder.
\newblock Physics of Fluids \textbf{14}(2), 628--637 (2002)

\end{thebibliography}

\end{document}